\documentclass[preprint,12pt]{elsarticle}

\usepackage{amssymb}
\usepackage{float}
\usepackage{amsmath}
\usepackage{booktabs}
\usepackage{multirow}
\usepackage{ulem} 

\usepackage{url}
\usepackage{hyperref}
\usepackage[hyphenbreaks]{breakurl}

\journal{Annals of Nuclear Energy}

\begin{document}

\begin{frontmatter}



\title{Validating Automated Resonance Evaluation with Synthetic Data}


\author[inst1]{Oleksii Zivenko}
\author[inst1]{Noah A. W. Walton}
\author[inst1]{William Fritsch}
\author[inst1]{Jacob Forbes}
\author[inst1]{Amanda M. Lewis}
\author[inst1]{Aaron Clark}
\author[inst2]{Jesse M. Brown}
\author[inst1]{Vladimir Sobes}

\affiliation[inst1]{organization={Nuclear Engineering Department, University of Tennessee},
            city={Knoxville},
            postcode={37996}, 
            state={TN},
            country={USA}}

\affiliation[inst2]{organization={Nuclear Data Group, Oak Ridge National Laboratory},
        city={Oak Ridge},
        postcode={37830}, 
        state={TN},
        country={USA}}

\begin{abstract}
The integrity and precision of nuclear data are crucial for a broad spectrum of applications, from national security and nuclear reactor design to medical diagnostics, where the associated uncertainties can significantly impact outcomes. A substantial portion of uncertainty in nuclear data originates from the subjective biases in the evaluation process, a crucial phase in the nuclear data production pipeline. Recent advancements indicate that automation of certain routines can mitigate these biases, thereby standardizing the evaluation process, reducing uncertainty and enhancing reproducibility. This article contributes to developing a framework for automated evaluation techniques testing, emphasizing automated fitting methods that do not require the user to provide any prior information. This approach simplifies the process and reduces the manual effort needed in the initial evaluation stage. It highlights the capability of the framework to validate and optimize subroutines, targeting the performance analysis and optimization of the fitting procedure using high-fidelity synthetic data (labeled experimental data) and the concept of a fully controlled computational experiment. 
An error metric is introduced to provide a clear and intuitive measure of the fitting quality by quantifying the estimate's accuracy and performance across the specified energy. This metric sets a scale for comparison and optimization of routines or hyperparameter selection, improving the entire evaluation process methodology and increasing reproducibility and objectivity.

\end{abstract}


\begin{highlights}
\item  Discusses the important question of how to measure evaluation quality, proposing a metric to estimate fitting accuracy and performance, making the nuclear data evaluation process more streamlined, less subjective, and more systematic.
\item Introduces a framework for benchmarking and optimizing automated evaluation instruments using high-fidelity synthetic data.
\item Demonstrates the concept and capabilities of the framework through an example: benchmarking and hyperparameter selection for a prior-free automated resonance identification subroutine. 

\end{highlights}


\end{frontmatter}


\section{Introduction}
\label{sec:introduction}

Nuclear Data (ND) is crucial in fields like national security, nuclear power, medical diagnostics, and scientific research. With goals to significantly increase nuclear energy capacity  \cite{tripleNEprodDOE2023COP28, NEA2023COP28} there's a push for new reactor designs and better performance of current ones. As we explore novel reactor designs with little existing data \cite{ND_adv_react}, the demand for precise ND is higher than ever. For example, in designing new nuclear reactors, studies \cite{CisnerosPresentation, Tahara2024} showed that the choice of evaluated ND for $^{35}$Cl could change a reactor's key safety measure, criticality, by about 2200 pcm. This difference makes a reliable core design difficult and strongly affects important core characteristics such as the transuranic elements transmutation rate, conversion ratio, etc. This highlights how crucial accurate ND is for the safety and design of nuclear reactors. Similar significance can be shown for other applications where accuracy ensures safety, boosts efficiency and drives innovation.

The generation of ND is a collaborative and extensive process, engaging the global nuclear data science community, including scientists, computational tool developers, and experimental facilities. This collective effort is often described as the ND production pipeline, which involves a series of steps: measurements, compilation, evaluation, processing, validation, and application \cite{Current_ND_needs_Kolos_2022}. The duration of this process, typically ranging from 5 to 15 years \cite{mccutchan_nd,Brown2021_pipeline_pres}, reflects the intricate nature of integrating experimental data with theoretical models and validation checks.  

Within this complex pipeline, the evaluation stage is particularly critical, relying heavily on expert judgment and thus presenting challenges in terms of reproducibility. The Working Party on International Nuclear Data Evaluation Cooperation (WPEC), subgroup 49, emphasizes the need for reproducibility in the ND evaluation process \cite{SG_49_WPEC}. 

Automation stands out as a promising solution to enhance this process - to minimize manual work and bias, improving the accuracy and reproducibility of ND evaluation. This has led to the incorporation of machine learning (ML) throughout the evaluation process, with examples of its application detailed in recent studies \cite{Valdez_MLAE, TALOU_sense_2021108568, schnabe_bayes_N}. One of the hurdles in applying ML is the scarcity of labeled data, a gap effectively bridged by the use of synthetic data (SD), which closely replicates real experimental data.

This article highlights the use of synthetic data in improving the ND evaluation process, particularly enabling hypotheses checks, testing, and training of ML-driven tools. An example is the hyperparameter optimization for an automated resonance identification subroutine \cite{walton2024_ARI}, which utilizes SD produced by established methods \cite{WALTON_Brown_phys_inf2024}. In this paper, the automated resonance identification subroutine, as part of the ATARI framework \cite{ATARI_repo}, is referred to as ``ARIS'' for brevity.

By leveraging SD, which allows for comparison against a ``true solution'' in a fully controlled virtual experiment, this approach not only highlights SD's crucial role in improving ND evaluations but also its effectiveness in enabling thorough and cost-efficient theoretical analyses and validation. This capability to conduct sensitivity studies, explore different methodologies, optimize future differential experiments and answer ``what if'' questions by testing various hypotheses is invaluable for practical applications and theoretical exploration in the ND domain.

It is important to acknowledge that the US ND community is currently navigating through challenges such as an aging workforce coupled with an increasing demand for new ND \cite{nsac_nd2023, Current_ND_needs_Kolos_2022, WANDA_2020_report, Bernstein_Fut_NDN_2019}, noting the importance of accurate ND for Gen. IV reactor design \cite{gen4_react_ND}.  Automation can reduce the evaluators' workload by streamlining repetitive tasks. Therefore, experienced evaluators can focus on mentoring and collaborating on new automated tools multiplying the effect, fostering knowledge transfer, and ensuring the sustainability and growth of the ND community. 

\section{Background}

Conventionally, the evaluation process in the resolved resonance region (RRR) relies heavily on popular codes like SAMMY \cite{sammy_man} or REFIT \cite{REFIT_MANUAL_Moxon2010} which use Generalized Least-Squares Fitting routines derived from a linearized version of Bayes Equations. However, evaluators face significant challenges in establishing reliable prior estimates of all resonance parameters - this step is crucial to establish local convexity and linear regime that will ensure accurate parameter estimation and fitting convergence. The vast number of resonances in typical evaluations makes this process tedious, time-consuming, and hardly reproducible \cite{Brown_ML_classif_NR}. Experts manually create priors, assign quantum numbers and vary resonance parameters to minimize $\chi^2$   striking a balance between a good fit and matching assumed distributions of resonance parameters \cite{Nobre_classif_2023} \footnote{Here, $\chi^2$ can be understood as a measure of the difference between observed and expected data values. In a linear model with normally distributed noise, the maximum likelihood estimate of parameters $P$ is obtained by minimizing the squared Mahalanobis distance between the model's predictions and the observations. This squared distance follows a $\chi^2$ distribution with degrees of freedom equal to the number of estimated parameters and serves as the goodness of fit.}. Consequently, this part of the evaluation is subjective and heavily influenced by expert judgments, hardly reproducible, which leads to significant amounts of unquantified uncertainty in the final evaluation. 

To streamline resonance parameter estimation and significantly reduce manual effort, an algorithm has been developed that eliminates the need to directly provide prior data about resonance locations and parameters \cite{walton2024_ARI}.  The algorithm employs a two-step approach: 1) beginning with the creation of a redundant feature bank for regression, addressing the problem's non-convex nature, and 2) applying backward stepwise selection to iteratively remove excess resonances, aiming to simplify the model and reduce over-fitting. This process, guided by domain knowledge, aims to optimize model complexity (number of resonances), ultimately minimizing the squared Mahalanobis distance ($\chi^2$). This routine is developed around SAMMY \cite{sammy_man} and uses the description of the experiments with all necessary experimental corrections provided (Doppler broadening,  resolution functions, etc.). The developed algorithm outputs a set of models and uses no prior information about location or parameters of resonances.

Selecting model complexity is not straightforward because the $\chi^2$ measure of fit will monotonically increase as model complexity decreases (in an ideal scenario where all possible models are explored). At the same time, a subset of candidate models can have very close values of $\chi^2$ \cite{walton2024_ARI}. While the $\chi^2$ as an objective function is used for direct optimization and serves as a goodness of fit measure, using it directly for model selection is a nontrivial task. In this case, several information-theoretic criteria can be used to strike a balance between the goodness of fit and model complexity for the observed data. Popular criteria for maximum likelihood or least squares regression problems are the Akaike Information Criterion (AIC) \cite{akaike1983information} and Bayesian Information Criterion (BIC) \cite{BIC_1}, which are standard outputs for each set of models.  Those criteria measure the balance between fitting quality against model complexity and have different penalty terms (BIC penalizes complex models more heavily than AIC). AIC/BIC values or corresponding $\chi^2$ values can be used for model selection directly, and there are well-established recommendations and strategies on that \cite{burnham2002model, Ding_Model_sel_tech}. 

However, practice in our RRR application shows that relying solely on generalized theoretical recommendations on thresholds $\Delta_{\chi^2}$ or $\Delta_{AIC}$ for model selection, on average, leads to models with an overestimated number of resonances (parameters) compared to the true model (overfitting). In other words, for our application these methods tend to favor more complex models that fit the data better but do not necessarily reflect the underlying physical reality, interpreting and fitting noise in the data as real resonances. For accurate calculation of penalized likelihood (AIC/BIC), one needs to provide a reasonable number of effective degrees of freedom, which is problematic for nonlinear models with highly correlated parameters \cite{andrae_chi2}. Additionally, the assumption about the normality of the residual distribution is often violated for our models. These issues lead to inaccurate penalty terms in AIC/BIC calculations, resulting in improper model selection when applying recommended thresholds.

In this case, criteria and corresponding thresholds for model selection serve as a hyperparameter of a fitting routine that needs to be optimized (selected). Using SD for this task allows the calibration of these parameters by evaluating how different values affect routine performance against known true values to minimize prediction errors. The methodology section will detail performance metrics and optimization strategies.

\section{Methodology}

\subsection{Synthetic data utilization for benchmarking of the subroutines and hypotheses testing }

\begin{figure}[H]
    \centering
    \includegraphics[width = \textwidth]{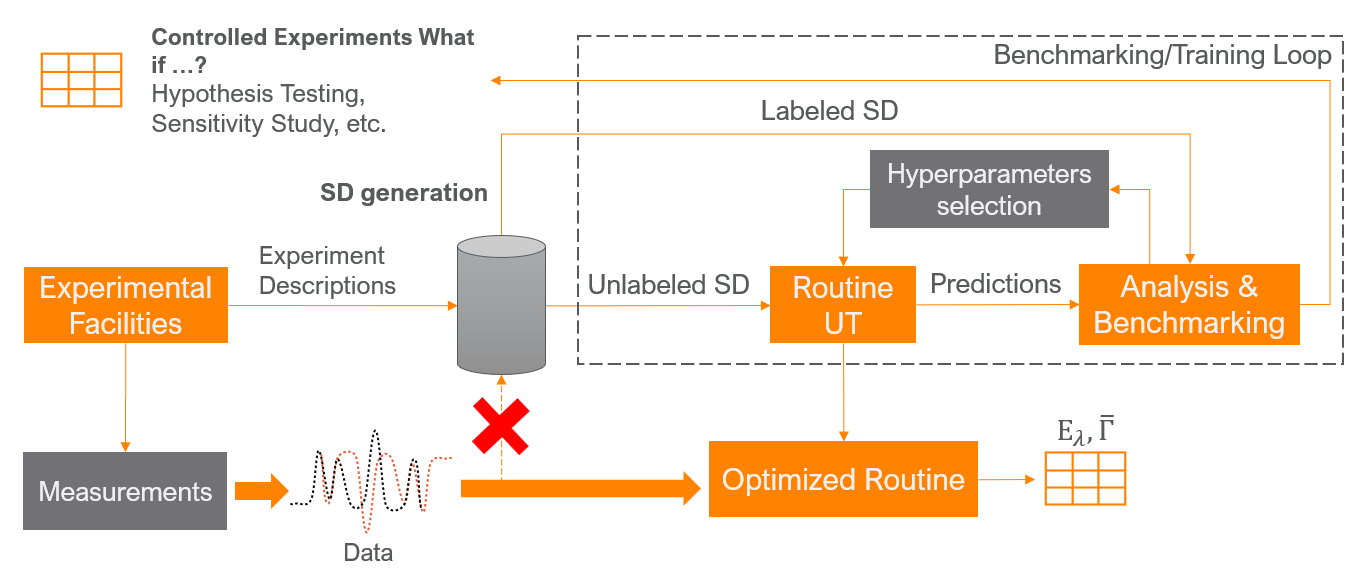}
    \caption{Utilization of synthetic data for assessment and optimization of routines under test and controlled computational experiments}
    \label{fig:training_loop}
\end{figure}

The framework under development utilizes high-fidelity SD and the concept of supervised learning \cite{HastieTibshiraniFriedman2009, MohriRostamizadehTalwalkar2012_FOML} having access to the assumed true solution \cite{CSEWG_2023_repr}. It also allows the introduction of various metrics for testing and consequent optimization of routines of interest by integrating those metrics into the assessment pipeline. Fig. \ref{fig:training_loop} illustrates this approach using the example of the optimization of the routine under test (UT).

Measurement data come from different experimental facilities accompanied by rigorously documented descriptions of corresponding experiments - all necessary data that fully characterize the experiment (data about experimental samples, conditions, processing techniques applied to the corresponding observables and uncertainties, etc.). In the next step, the dataset for testing and/or benchmarking the routine under test is generated. It consists of simulated, high-fidelity experimental observables for each corresponding experiment that mimic real experimental data provided for evaluation. It's worth noting that the generated SD is statistically indistinguishable from real experimental observed data, allowing the learned hyperparameters to be extrapolated to real, unseen data. The generation process is physically informed to reflect the experiment flow and add appropriate noise to the radiation detection signals so the synthetic observables are statistically similar to real ones. Details on the synthetic data generation methodology can be found in \cite{WALTON_Brown_phys_inf2024}. 

The presented training loop naturally opens the way for different studies that can be done based on unique modeling ability and access to effectively unlimited, labeled datasets. Thus, utilizing the idea of controlled experiments, one can perform an in-depth analysis of various factors that may affect routine performance and responsiveness to known variables. At the same time, the exploration of the routine behavior under different hypothetical scenarios can be analyzed (incorrect data covariance matrices, different initial assumptions, modification of variables with unknown data about correlations, etc.), and corresponding recommendations can be produced. As another potential application - the production of reliable data on uncertainty (covariances) using specific fitting techniques and assumptions about the models and sets of corresponding synthetic data can be performed, similar to the process described in  \cite{Rochman2023_covariances_variation_models}.  
This also includes the potential to enhance experiment design to maximize value from the measurement process and generate reliable conclusions based on statistical studies and benchmarking.

\begin{figure}[H]
    \centering
    \includegraphics[width = \textwidth]{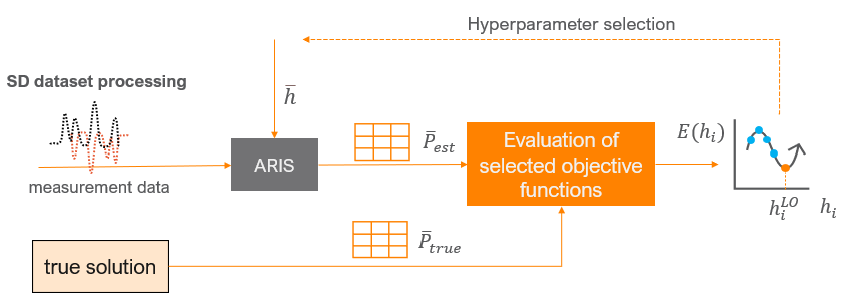}
    \caption{Selection of hyperparameters for the automated resonance identification routine, treated as a "black box", involves using access to the true solution and selected metrics of interest}
    \label{fig:ARI_hp_sel}
\end{figure}

In this analysis, the automated resonance identification subroutine is treated as a black-box tool, implying that the intricacies and specific functions or interactions of its hyperparameters remain opaque to us. For an illustrative explanation, see Fig. \ref{fig:ARI_hp_sel}.

ARIS processes SD and produces estimates for resonance parameters \( \overline{P}_{\text{EST}} \). The ARIS is controlled by a vector of hyperparameters \( \overline{h} = [h_1, h_2, \ldots, h_n] \), where each hyperparameter is characterized by its specific range and intended role. Having access to the values of underlying true resonance parameters \( \overline{P}_{\text{TRUE}} \) and designing a set of objective functions \( \overline{E}_{\text{OBJ}} \), we can perform an in-depth analysis to estimate the automated resonances identification routine's accuracy and sensitivity for any set of hyperparameters under test and select  $h^{\text{LO}}$ - set of locally-optimal parameters. We refer to the selected hyperparameters as `locally-optimal" because we assume their effectiveness is tailored to specific SD that mimics experimental observables for certain conditions. Furthermore, the estimation of these parameters is indirectly constrained by the testing scheme and other variables, which include a selected grid and other influential variables not categorized as hyperparameters. These factors collectively limit the generalizability of estimation, indicating that while these parameters are the best under our current test conditions, they may not be universally optimal across different or more varied datasets. This term "sub-optimal" thus emphasizes the conditional effectiveness of our hyperparameters, acknowledging that further adjustments might be necessary to optimize performance in other contexts or with broader data variabilities.

\subsection{Idealized Cross Section Evaluation Metric}

We pose that it is difficult to select a single metric to compare the quality of a resonance evaluation bound for the ENDF/B library (general purposes evaluated nuclear data library).  Perhaps, without access to a \emph{true} cross section associated with a synthetic data set, the nuclear data community has not yet had the opportunity to hypothesize what such a metric would look like.  For example, consider two resonance evaluations, \emph{ladder A} and \emph{ladder B}. Suppose \emph{ladder A} has one extra resonance compared to \emph{ladder B}, while \emph{ladder B} has the same number of resonances as the underlying \emph{true ladder}. Nevertheless, \emph{ladder A}  resulting cross section is closer to the \emph{true} in the evaluated cross section space. Which evaluation is better?  Which one should we, as a community, prefer?  What does it even mean, quantitatively, to be closer in the evaluated cross section?  These are the types of questions that synthetic data and/or \textit{access to the true solution} allow us to begin to discuss.

From the understanding of the authors, the uses of the data in the resolved resonance region in the ENDF/B, there are two separate things that matter: 1) the evaluated cross section (the Doppler Broadened cross section at 293K and above) and 2) the average resonance parameters.  In the second item, it is important to point out that, as far as we understand, in current practice, the individual resonance parameters are not used anywhere other than to reconstruct the energy-dependent cross section (and, at times, angular distributions of elastic scattering) and to estimate the values of the average parameters to inform the unresolved resonance region evaluations and optical model parameters.

Now, we turn to the question of what is an appropriate way to measure agreement in cross-section-space.  Suppose that we have an \emph{imperfect} automated evaluation that produces the best estimate of the cross section (with some error), and we also have a \emph{perfect} uncertainty quantification routine that produces statistically consistent error estimates for the evaluated cross section (e.g. represented by a covariance matrix in ENDF File 33 assuming a multivariate normal distribution).  As a simplified example, the cross section evaluation routine will not always get the \emph{true} cross section exactly right but the uncertainty evaluation routine will be such that the \emph{true} cross section will be within one standard deviation of the evaluated cross section 68\% of the time as predicted by normal statistics.  So, we propose that under these idealized circumstances, as a community, we should prefer the resonance evaluation with a smaller uncertainty estimate.  That is to say, that, really, all we should care about is not the evaluated cross section but rather the upper and lower limits of where we can confidently (and quantitatively) say that the \emph{true} cross section lies.

However, this does not fully resolve the ambiguity but, rather, pushes it further down the line.  From, how does one measure agreement in cross-section-space the question becomes how does one compare the relative size of cross section uncertainty estimates?  Let's narrow it down to an idealized case of multi-variate normal distributions parameterized by cross section covariance matrices.  We will consider a single cross section covariance matrix produced on a dense energy grid, covering all partial reactions (e.g. elastic scattering, capture, fission, etc.), with cross correlations. This simplification allows us to reduce the problem to comparing the \emph{sizes} of covariance matrices, which have several well-established norms (e.g. the Frobenius-norm, the nuclear-norm, and the 2-norm).  

Where are cross section covariances used and where do they impact the users of the ENDF/B library?  It is in the propagated nuclear data uncertainty on the applications.  This is calculated either through Total Monte Carlo methods \cite{TMC_Koning2011, TMC_master_thesis_Vaara2022} or as a linear propagation of uncertainty (which becomes a good approximation when the nuclear data uncertainty is small). Linear propagation of uncertainty is computed as:

\begin{equation}
    (\delta k)^2 = \frac{\partial k}{\partial \sigma}(\delta \sigma)^2 \frac{\partial k}{\partial \sigma},
    \label{eq: sandwich rule}
\end{equation}
\noindent
where, $k$, stands for a generic application response (used to suggest the often-used k-eigenvalue of critical systems), and $(\delta \sigma)^2$ represents the cross section covariance matrix for all reactions (including cross-correlations) evaluated on a dense energy grid of incident neutron energies.

Which application should we consider?  There is a strong push in the nuclear data community for ENDF/B library to be application agnostic \cite{ENDF8_BROWN, neudecker2021covariance}, which would seemingly put a dead-end to this line of reasoning.  However, the 2-norm of a covariance matrix, ($||C||_2$), has a very convenient interpretation deduced from the following identity:

\begin{equation}
    \max_{x} \frac{||Cx||_2}{||x||_2} = ||C||_2
    \label{eq: 2 norm}
\end{equation}

If we choose to accept the resonance evaluation with the smallest consistently-evaluated cross section covariance matrix measured by the 2-norm, then we can interpret this as minimizing the maximum possible linearly-propagated error to any application from nuclear data.  This is seen by interpreting the $x$ in Equation~\ref{eq: 2 norm} as the sensitivity, $\partial k/\partial \sigma$ in Equation~\ref{eq: sandwich rule}.  This choice has a tangible connection to applications while maintaining the ENDF/B-preferred neutrality of the choice of application system.  

\subsubsection{A Softened Cross Section Evaluation Metric}

In practice, however, optimizing a fitting procedure to maximize the performance on a metric based on the maximum eigenvalue of the cross section covariance matrix (a restatement of the matrix 2-norm) is difficult.  Therefore, we chose to work with the Frobenius norm of the covariance matrix and consider it a \emph{softened} version of the 2-norm.  The Frobenius norm of a matrix is equivalently represented as the sum of the squares of the eigenvalues.  If the covariance matrix has only one dominant eigenvalue then the Frobenius norm is a good upper bound of the 2-norm.  For our work, the Frobenius norm is also convenient to compute because it is equivalently the sum of the squares of all of the entries of the covariance matrix, making it easy to estimate in this work.  In addition, if we initially neglect cross energy and cross reaction correlations as well, this simply reduces to the squared error summed over all energies and all reactions. For the specific example of benchmarking and selecting hyperparameters for the ARIS, the metric that measures accuracy in the cross-section space for a given reaction type can be approximated as follows: 

\begin{equation}
\label{eq:CS_error_one_react}
\varepsilon_{\sigma_{\text{react}}} = \frac{{ \int_{E_{\text {min}}}^{E_{\text {max}}} (\sigma_{\text {fit}}^{\text {react}} - \sigma_{\text {true}}^{\text {react}})^2 \, dE}}{\int_{E_{\text {min}}}^{E_{\text {max}}}dE},
\end{equation}
\noindent
where $\sigma_{\text {fit}}^{\text {react}}$ is a cross section calculated using estimated parameters of the resonances (from the fitting routine under test) and $\sigma_{\text {true}}^{\text {react}}$ is the cross section for true resonances parameters for selected reaction type; $E_{\text{min}}$ and $E_{\text{max}}$ are the bounds of energy region used for analysis.

\begin{figure}[H]
    \centering
    \includegraphics[width=\textwidth]{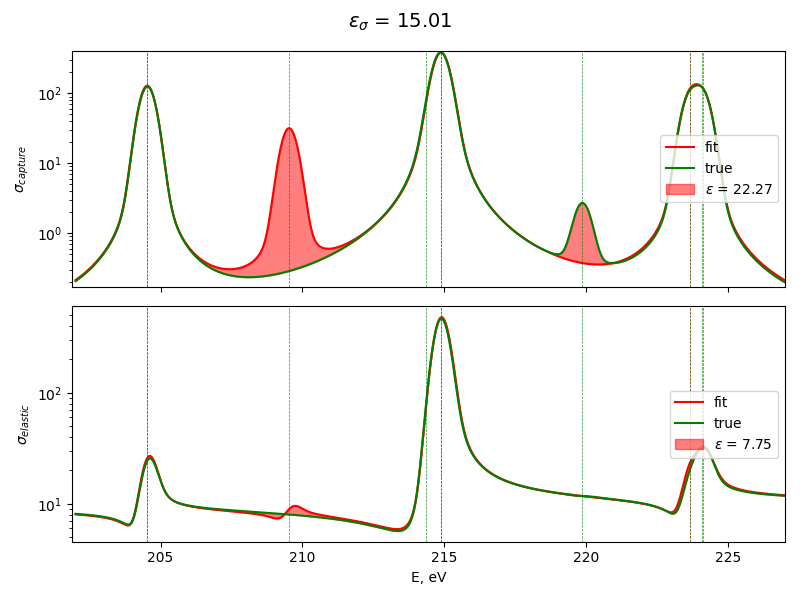}
    \caption{Proposed error metric $\varepsilon_{\sigma}$ provides an aggregate measure of the fitting accuracy for specific reaction type and/or across the entire spectrum of reactions within a given energy region; and can be calculated as an integral of the squared difference between estimated and true cross sections (taking into account Doppler broadening at reference temperature)}
    \label{fig:error_metric_0}
\end{figure}

This metric can be useful for practical applications. It can be easily interpreted as the mean squared deviation of the fitted cross-section from the true cross-section over a specified energy range (see illustration of the principle shown in Fig. \ref{fig:error_metric_0}). Green vertical lines show the positions of the true resonances and red - positions of the resonances for the estimated resonance ladder. 

As illustrated in Fig. \ref{fig:error_metric_0}, discrepancies such as missed resonances (e.g., near 215 or 220 eV) or ``fake'' resonances - erroneously inserted resonances (near 209 eV) that appear in the estimation without corresponding to any actual resonance. The integration over energy range provides a single measure of overall fit quality for a given energy range. 

To apply the error metric to the real world, where we deal with a discrete form suitable for practical computation, the integral must be replaced with a sum over a finely discretized energy grid. To combine the error metrics across all reactions into one comprehensive metric, we sum the individual squared errors for each reaction and normalize by the number of energy points. 

\begin{equation}
\label{eq:CS_error_all_react_sum}
\varepsilon_{\sigma} = \frac{\sum_{\text react}  { \sum_{i=1}^{N_{\text E}} (\sigma_{\text {fit}}^{\text {react}}(E_{i}) - \sigma_{\text {true}}^{\text {react}}(E_{i}))^2}}{N_{\text {react}}N_{\text E}}
\end{equation}

In equation \ref{eq:CS_error_all_react_sum}, $E_{i}$ represents the $i$-th energy point on the grid, $i$ = 1 to $N_{E}$, where $N_{E}$ is the total number of points within the selected energy range [$E_{\text{min}}$, $E_{\text{max}}$], $N_{\text react}$ - number of considered reactions. The energy grid's density is defined by a default of 100 points per eV, ensuring that the summation closely approximates the intended integral. Here, $\sigma_{\text {fit}}^{\text {react}}(E_{i})$ and $\sigma_{\text {true}}^{\text {react}}(E_{i})$ are the cross-section values at $E_{i}$ for the fitted and true parameters, respectively, for each reaction type under consideration calculated at some reference temperature (e.g. 300 K).

This consolidated error metric can be interpreted as a mean squared error in cross-section space at each point of energy per reaction, calculated as the mean error along the incident neutron energy spectrum. It provides an aggregate measure of the fitting routine's accuracy across the entire spectrum of reactions within a given energy region, thus offering a single number that characterizes the routine’s performance for all reactions of interest.

It's also useful to have an additional metric that characterizes the solution in resonance parameters space and is sensitive to the widths, the locations, and the spin group assignments. It is particularly relevant as it enhances our understanding of how individual resonance characteristics influence the aggregated behaviors captured in average parameters, which are crucial to the community.  For this, we use strength functions as a basis for the new metric, which for the partial waves $l$ is defined as \cite{Neutron_xs_Resonance_parameters_Mughabghab}: 

\begin{equation}
\label{eq:SF_Gn}
S^{l} = \frac{1}{(2l+1)\Delta E} \sum_{j}{g_{j} \gamma^2_{nj}},
\end{equation}
\noindent
where the summation is carried over $N$ resonances in energy interval $\Delta E$, $g_{j}$ is the spin statistical weight factor for angular momentum $J$ and  $\gamma_{nj}^{2}$ - reduced neutron width. \footnote{To ensure clarity in notations, we adhere to the terminology used in the SAMMY manual: reduced width amplitude is denoted as \( \gamma \), reduced width as \( \gamma^2 \), and partial width as \( \Gamma_{\lambda} = 2 P(E_{\lambda}) \gamma_{\lambda}^2 \).}

Eq. \ref{eq:SF_Gn} gives strength function value for a given resonance ladder and energy range, taking the squared difference and summing for all spin groups we can define:
\begin{equation}
\label{eq:SF_Gn_sq_difference_all_sg}
\varepsilon_{\text {SF}} = ({S^{l}_{\text{fit}} - S^{l}_{\text{true}}})^2,
\end{equation}
\noindent
where $S^{l}_{fit}$, $S^{l}_{true}$ - strength function values calculated using eq. \ref{eq:SF_Gn} for estimated and true resonance parameters.

To characterize the quality of fitting in resonance parameters space and incorporate the information on spin group assignment, we define a metric $\varepsilon_{\text{PSF}}$ similar to eq. \ref{eq:SF_Gn_sq_difference_all_sg} , with the only difference that partial strength functions are used (calculated separately for each spin group considered):

\begin{equation}
\label{eq:SF_Gn_sq_difference_by_sg}
\varepsilon_{\text {PSF}} = \sum_{J^{\pi}}  ({S^{J^{\pi}}_{\text{fit}} - S^{J^{\pi}}_{\text{true}}})^2,
\end{equation}
\noindent
where $S^{J_{\pi}}_{\text{fit}}$, $S^{J_{\pi}}_{\text{true}}$ - strength functions for estimated and true ladder respectively calculated using resonances of each spin group separately (for each $J^{\pi}$).

The error metric is calculated as a sum of squared differences in partial strength functions to capture the difference in estimated and true resonance parameters over the energy range of interest. 

\begin{figure}[H]
    \centering
    \includegraphics[width=\textwidth]{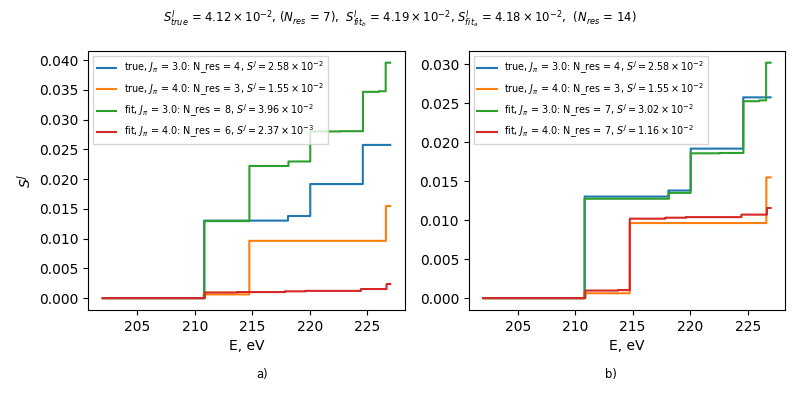}
    \caption{Illustrating the introduced error metric calculated using strength functions: 
    a) - results for a case with incorrect resonance spin groups assigned for part of resonances; 
    b) - results after reassignment of the spin group for one resonance}
    \label{fig:error_metric_1}
\end{figure}

The metric calculated using eq. \ref{eq:SF_Gn_sq_difference_by_sg} allows us to be more sensitive for parameter differences and specific error types: incorrect spin group assignment (significant number of missed/fake resonances). Fig. \ref{fig:error_metric_1} exemplifies the underlying principle: where the true ladder has 7 resonances of two spin groups ($J^{\pi}$ = $3.0^+$ and $J^{\pi}$ = $4.0^+$) in a given energy region, while the estimated ladder has 14 resonances. A straightforward calculation of strength functions for both ladders gives approximately equal values $S^{l}_{\text{true}} \approx S^{l}_{\text{fit}}$ (calculation results for presented example are given in table \ref{tab:error_metrics_comp}).

Notably, all of the true resonances were correctly identified for a given case, but the spin group assignment for part of the resonances was incorrect. Specifically, for a resonance at 214.75 eV, the spin group was initially assigned incorrectly - see Fig. \ref{fig:error_metric_1} a). For this resonance, the spin group was flipped, and the model was then refitted to the same data with the spin group reassigned - Fig. \ref{fig:error_metric_1} b). The error metrics before and after the spin group reassignment (SGR) are presented in Table \ref{tab:error_metrics_comp}. This adjustment highlights the impact of correct spin group assignments on the metrics $\varepsilon_{\text{SF}}$ and $\varepsilon_{\text {PSF}}$, underscoring their different sensitivity to errors in spin group classification. 

\begin{table}[H]
    \centering
    \begin{tabular}{lrrr}
    \toprule
         &  $\varepsilon_{\sigma}$, $\text{barns}^2$& $\varepsilon_{\text{SF}} \times 10^{-7}$& $\varepsilon_{\text{PSF}} \times 10^{-4}$\\
    \midrule
         Initial spin group assignment&  13.16& 4.94&3.63 \\
         After SGR \& refit&  6.42&2.72&0.35 \\
    \bottomrule
    \end{tabular}
    \caption{Comparison of error metrics for the example case processed}
    \label{tab:error_metrics_comp}
\end{table}
 
The metrics introduced in this section serve as an illustrative example. The choice of metrics is highly flexible and can be tailored to the specific objectives and nuances of the analysis. The fundamental point emphasized here is the significant advantage provided by having access to a true solution. This, along with employing multiple metrics, empowers researchers to conduct more comprehensive analyses and evaluations.

\subsection{Hyperparameter selection. Testing data}

This section shows a testing scheme for the hyperparameter optimization and assessment of ARIS utilizing different assumptions and asking ``what if?'' questions. It must be clarified that this paper does not delve into the intricacies of each parameter within ARIS. The primary objective is to demonstrate the assessment of the subroutine and integrating error metrics of interest, refining hyperparameters using synthetic data. Nonetheless, a short explanation is provided to foster an understanding of the subroutine's functionality, even as it is treated as a computational ``black box'' for the purposes of this evaluation.

The essence of this methodology lies in analyzing the routine's efficiency through predefined metrics, aiming to choose the hyperparameters based on the assessment results. The ARIS is controlled by a set of parameters, each contributing uniquely to the operational flow. These parameters are stratified into three main categories, distinguished by their influence and inherent properties.

The first group of parameters stands for parameters of the initial feature bank (IFB), such as the density of the feature bank (or the size of the initial feature bank), starting values for all of the parameters of the feature bank, presence of the resonances of all spin groups for isotope under investigation, presence of the fixed resonances outside the window, etc. The second group of parameters controls the behavior of the core fitting routine (step sizes, batching for the data and/or model parameters, etc.). 
The third group controls the backward stepwise regression phase, wherein the process iteratively eliminates resonances based on some predefined weighting criterion. For an exhaustive delineation of the parameters of the automated resonance identification tool, the reader is directed to \cite{walton2024_ARI}.

This study modulated two parameters to investigate the sensitivity: $N_{\text{IFB}}$, denoting the resonance count per spin group within the IFB for a given energy region, and $\Delta\chi^2$, the threshold used during the backward stepwise regression phase. This dual parameter variation serves to underscore the differential impact of parameter types within ARIS and to exhibit the practicality of manipulating multiple parameters simultaneously.

We noted that increasing $N_{\text{IFB}}$, representing the ``density'' of the IFB, consistently enhances the accuracy of the resonance identification by providing a richer set of initial features for model construction. However, this comes at the cost of significantly increased processing time due to the more extensive feature set. Given this predictable impact on performance and computational load, $N_{\text{IFB}}$ diverges from the traditional definition of a hyperparameter because its effect on the subroutine's efficiency is foreseeable. It's also worth mentioning that in real-life applications, one is always limited by the amount of data to fit in the energy region of interest. Hence, $N_{\text{IFB}}$ can also be limited and even initially chosen using the data of theoretical calculations to overcome the convexity problem. For results presented later in this paper, after testing different values of feature bank densities for selected energy regions, we choose $N_{\text{IFB}} = 50$  resonances per spin group to balance between accuracy and processing time. That corresponds to the distance of 0.5 eV between two neighboring candidate resonances of the same spin group for the selected energy range. 

Conversely, the parameter $\Delta\chi^2$ operates as a more conventional hyperparameter. It functions as a threshold in the backward stepwise regression phase, defining when the regression process should terminate. The optimal value of $\Delta\chi^2$ is not readily apparent and requires tuning to strike an optimal balance between model complexity (number of resonances) and accuracy. The selection of the threshold $\Delta\chi^2$ involves a search across a range of potential values to identify a value that maximizes the ARIS performance under specific conditions and data characteristics (e.g. energy range and applied density of the initial feature bank or other values that affect performance). 

We emphasize that this value is connected with the nature of the data and noise present in the experimental datasets. In theory, the nature of this threshold can be explained as a balance between the model complexity (the number of resonances decreased with each step) and the absolute change of the measure of goodness of fit - $\Delta\chi^2$. This assumes that we always find a minimum of $\chi^2$ for a given model. However, we are limited by other factors that affect initial resonance identification and/or fitting quality. So, this threshold must be selected, considering a complex relationship with other parameters that can affect the behavior and performance of the routine under test. 

Multiple datasets were generated for training, hypotheses testing, and validation. Their characteristics are presented in Table \ref{tab:gen_datasets_table}. The purpose and differences of these datasets are described below. 

\begin{table}[H]
\centering
\begin{tabular}{@{}cccp{4cm}l@{}} 
\toprule
\textbf{Name} & \textbf{Purpose}& \textbf{$N_{\text{SG}}$} & \textbf{Distortion}& \textbf{Study}\\
\midrule
A& TR  & 1 & None & 1.1\\
B& V   & 1 & None & 1.1 \\
A& WI& 1 & DCM  & 1.2 \\
A& WI& 1 & incorrect $ \langle \gamma_{\gamma}^2 \rangle $ & 1.3 \\
C& TR  & 2 & None & 2.1 \\
D& V   & 2 & None & 2.1 \\
C& WI& 2 & DCM  & 2.2 \\
\end{tabular}
\caption{Datasets generated for training (TR), validation (V) and ``what if?'' tests (WI) exampled by distorted DCM or assumption about average resonance parameters}
\label{tab:gen_datasets_table}
\end{table}

Here, we define a dataset as a large number of synthetically generated resonances and corresponding experimental observables. All datasets were generated using the methodology from \cite{WALTON_Brown_phys_inf2024}. The generative model for resonances was based on the Ta-181 isotope (averages determined from JEFF3.3 evaluation), while the generative model for the observables was based on the measurements made in  \cite{BrownThesis_2019} and \cite{Ta181_measurements_Brown}. 
The energy region of 202-227 eV was used for testing purposes to maintain consistency with the results presented in \cite{walton2024_ARI} where ARIS was applied to the real measurements data.

Each dataset consists of 500 independent cases for processing. For each case, three transmission measurements and two capture measurements across varying target thicknesses were generated. A full data covariance matrix (DCM) is sampled for the transmission data, while only diagonal values were generated for the capture data. This limitation is due to the nonvalidated methodology for generating high-fidelity uncertainty data for capture measurements, with the additional argument that it is common in older measurements to have data for diagonal uncertainties only. 

Each dataset serves a specific study estimating ARIS performance under different scenarios. Particularly, the performance of the subroutine for the next scenarios is assessed: 1) when ARIS is supplied with correct covariance data and correct values of average resonance parameters are known; 2) when only part of the covariance data is provided while the average resonance parameters are known 3) when full and correct covariance data is provided while incorrect assumption on the average resonance parameters is used. 

The performance of ARIS is compared for 2 groups of test scenarios: when data includes resonances of single or multiple spin groups, exploring the impact of correct spin group assignment. Thus the first study group estimates the performance of ARIS when the data contains resonances from a single spin group, thereby eliminating uncertainty related to correct spin group assignment. The simplest way to simulate this scenario is to generate observables that include resonances only of the first spin group, \( J^{\pi} = 3.0^+ \) (datasets with  \( N_{\text{SG}} = 1 \) in table~\ref{tab:gen_datasets_table}). The datasets of the first group (A, B) were generated with resonances from only one spin group $J^{\pi}=3.0^{+}$, while datasets (C, D)  contain data for both spin groups. Average parameters for resonances are shown in Table \ref{tab:avg_res_pars_by_sg_gen}.

Dataset A is used for the initial estimation of hyperparameter value (denoted as TR - training in Table~\ref{tab:gen_datasets_table}). For this test, ARIS is supplied with correct DCMs (full covariance matrices without any distortions), and correct values of average parameters are used during the fitting process. 

Dataset B serves as a validation dataset (marked V in the Purpose column of Table~\ref{tab:gen_datasets_table}) - to check if the performance on unseen data is consistent with the results obtained from Dataset A.

Studies 1.2 and 1.3 showcase two different ``What if?'' tests (denoted WI) for the scenario of a single spin group, utilizing the same observables as Study 1.1 (Dataset A). However, Study 1.2 supplies ARIS with misreported covariance data, and Study 1.3 supplies ARIS with incorrect values for average resonance parameters. Notably, an increased value of reduced capture width is assumed: \( \langle \gamma_{\gamma}^2 \rangle_{\text{assumed}}^{J^{\pi} = 3^+} = 35.2 \) meV.

Group 2 studies use data that includes resonances for both spin groups. The average resonance parameter values used for data generation are shown in Table~\ref{tab:avg_res_pars_by_sg_gen}.

\begin{table}[H]
\centering
\begin{tabular}{@{}ccc@{}}
\toprule
\textbf{}                               & \textbf{\( J^{\pi} = 3.0^+ \)} & \textbf{\( J^{\pi} = 4.0^+ \)} \\
\midrule
 $\langle D \rangle$, meV& 9003.0&8303.1\\
\( \langle \gamma_{\gamma}^2 \rangle \), meV & 32& 32\\
\( \langle \gamma_{n}^2 \rangle \), meV & 452.56                         & 332.24                         \\ \bottomrule
\end{tabular}
\caption{Average resonance parameter values used to synthesize experimental data for the studies of the second group}
\label{tab:avg_res_pars_by_sg_gen}
\end{table}

Study 2.1 uses dataset C and is directed to the initial performance estimation when ARIS is applied to the data with resonances of multiple spin groups. Dataset D serves for validation in this case. The results are then compared with those of study 1.1. Study 2.2 showcases the difference in performance if ARIS is supplied with incomplete covariance information, similar to Study 1.2.

To assess the performance for each processed dataset, both the average and median error metrics were calculated as follows:

\begin{equation}
\label{eq:dataset_avg_E}
\langle \varepsilon \rangle = \frac {1}{N_{\text{cases}}} \sum_{k=1}^{N_{\text{cases}}} E_{k}, \quad \text{Med}(\varepsilon) = \text{Med} \left\{\varepsilon_1, \varepsilon_2, ..., \varepsilon_{N_{\text{cases}}} \right\}
\end{equation}

In equation \ref{eq:dataset_avg_E}, $\varepsilon_{k}$ denotes the chosen metric of interest (either $\varepsilon_{\sigma}$ or $\varepsilon_{\text{PSF}}$) for the $k$-th case, with $k$ varying from 1 to $N_{\text{cases}}$, the total number of cases within the dataset.

The selection of the $\Delta\chi^2$ threshold is adjusted to minimize the average error $\langle \varepsilon \rangle$. Additionally, the median value, which provides an additional error measure, is calculated and reported for each processed dataset. 

To provide the information about the uncertainty associated with the calculated error metric value for the dataset the standard error of the mean is reported. To assess the uncertainty tied to the median of the same metric, the bootstrapping \cite{HastieTibshiraniFriedman2009} technique was used: recalculating the median value 1000 times for each  random subset. The uncertainty of the median is then reported as the standard deviation of these median values. 

\section{Results}
Hyperparameter selection, specifically the selection of the $\Delta\chi^2$ threshold, is demonstrated by minimizing the average error $\langle \varepsilon_{\sigma} \rangle$ calculated in cross-section space as described by Eq. \ref{eq:CS_error_all_react_sum}. Only this metric was utilized for selection, while all other relevant metrics are systematically presented in tabular form for clarity and comprehensive analysis. The selection of an appropriate objective function pertains to the specific requirements of each study and is beyond the scope of this article. It's also worth mentioning that all results presented in this article were produced using the following configuration of the ARIS and related limitations: correct average $\gamma_{\gamma}$ values were used as starting parameters for the IFB stage for all studies except Study 1.3; spin reshuffling option was not used for all stages of the automated resonance identification, during the intermediate pruning stage $\gamma_{\gamma}$ values for all resonances were fixed; solution selection mode from the subset - minimizing $\chi^2$  value (no prior information is used or any information about resonances outside the energy region under processing);  Reich-Moore approximation to the R-matrix equations was used \cite{ReichMooreApprox, sammy_man}. 

\subsection {Initial assessment and hyperparameter selection on the dataset with resonances of a single spin group}

\begin{figure}[H]
    \centering
    \includegraphics[width=\textwidth]{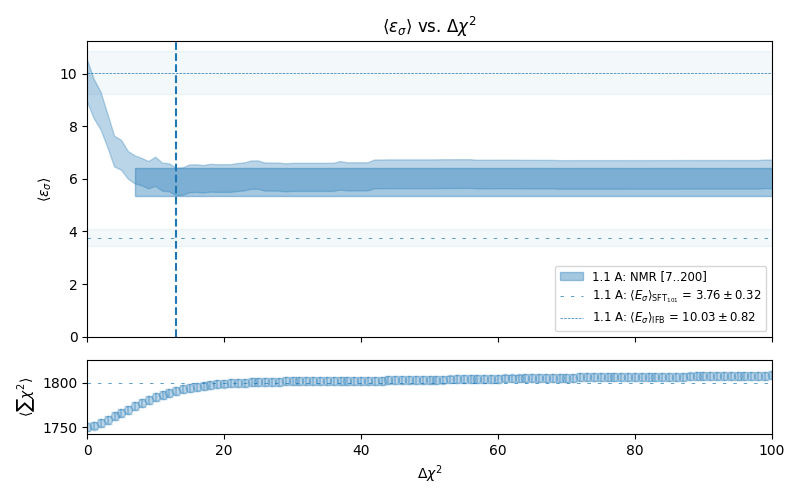}
\caption{Hyperparameter $\Delta\chi^2$ selection based on the error metric $\langle \varepsilon_{\sigma} \rangle$ (Study 1.1). Horizontal lines indicate reference values comparing optimized and ``best possible'' performance scenarios, including a ``starting from true'' solution and post-IFB stage performance. The vertical line marks the minimum of $\langle \varepsilon_{\sigma} \rangle$.}

    \label{fig:result_avg_training_ds}
\end{figure}

Results of the ARIS initial assessment using the data, which included resonances of a single spin group (dataset A), are presented in Table \ref{tab:res_1.1_A} and partially shown in the Fig. \ref{fig:result_avg_training_ds}.

\begin{table}[H]
\centering
\begin{tabular}{l|rr|r}
Metric& \multicolumn{2}{c|}{Estimated}              & Reference       \\
                                  & IFB stage        & Pruning Stage, $\Delta\chi^2_{\text{sel}}$ & SFT$_{101}$     \\ \hline
        $\langle \varepsilon_{\sigma} \rangle $ & $10.03 \pm 0.82$& $5.89 \pm 0.52$& $3.76 \pm 0.32$\\
        med($\varepsilon_{\sigma}$) & $2.67 \pm 0.39$& $1.6 \pm 0.16$& $0.91 \pm 0.11$\\
        $\langle \varepsilon_{\text{PSF}} \rangle $$\times 10^{9}$ & $1.32 \pm 0.11$& $1.88 \pm 0.21$& $1.91 \pm 0.2$\\
        med($\varepsilon_{\text{PSF}}) $$\times 10^{10}$ & $4.72 \pm 0.44$& $3.31 \pm 0.56$& $2.45 \pm 0.64$\\
        $\langle N_{\text{res}} \rangle $ & $22.03 \pm 0.19$& $3.28 \pm 0.07$& $2.79 \pm 0.04$\\
    \end{tabular}
    \caption{Estimated and reference error metrics (Results for Study 1.1, dataset A)}
\label{tab:res_1.1_A}
\end{table}

Two additional values were calculated to introduce the scale for comparison and show the impact of selected hyperparameter values. The first represents an estimate of the ``best possible average performance'' for the given dataset. This value is calculated assuming an ideal initial guess: precise positions and correct parameters of resonances (including data on spin group assignment), using the same technique to fit the model to the given data. It is denoted as ``start from true'' ($\text{SFT}_{101}$ - note, for the selected mode - the values of capture widths were fixed during the fitting process). The second reference value shows the result of the initial processing stage - initial features bank solve (denoted as IFB). For both reference values, the standard error of the calculated mean is reported, and corresponding regions are highlighted on the plots. On the figures, we additionally denote the \textit{Near Minimum Region} (NMR) as the range of hyperparameter values for which error metric value \( \langle \varepsilon \rangle \) lies within the $\varepsilon_{\text{min}} \pm \text{SD}_{\varepsilon_{\text{min}}} $ range, where  $\text{SD}_{\varepsilon_{\text{min}}}$ is an error in estimating the corresponding metric (mean or median estimation error). The data on $\langle \varepsilon_{\text {SF}} \rangle$ for  these studies are excluded from tables because correct values for the single spin group were passed for processing  $\left (  \langle \varepsilon_{\text {SF}} \rangle = \langle \varepsilon_{\text {PSF}}\rangle \right )$.

The Fig. \ref{fig:result_avg_training_ds} demonstrates the relationship between the error metric  $\langle \varepsilon_{\sigma} \rangle$ and various $\Delta\chi^2$ thresholds applied for processing of cases for dataset \textit{A}. 

The minimum value identified for $\langle \varepsilon_{\sigma} \rangle$ corresponds to $\Delta\chi^2_{\text {sel}}=13$, illustrating the impact of the threshold on this performance metric. It can be noted that there is a ``plateau'' with almost equal values for average error for the region of hyperparameter values highlighted as NMR. Additionally, the second subplot of the Fig. \ref{fig:result_avg_training_ds} shows the average value $\langle \sum{\chi^2} \rangle$ representing the sum of $\chi^2$ values for each fitted observable averaged by cases number of the dataset processed. 

This subplot also includes data on the value ${\langle \sum{\chi^2} \rangle}_{\text{SFT}_{101}}$, which represents the average sum of $\chi^2$ values calculated for each case of the dataset in ``start from true'' mode. The fact that after the IFB stage a large number of resonances are present in the fitting results and significant difference in goodness of fit measure values  ${\langle \sum{\chi^2} \rangle}_{\text{IFB}} << {\langle \sum{\chi^2} \rangle}_{\text{SFT}_{101}}$ proves that ARIS significantly overfits the data. 

\begin{figure}[H]
    \centering
    \includegraphics[width=\textwidth]{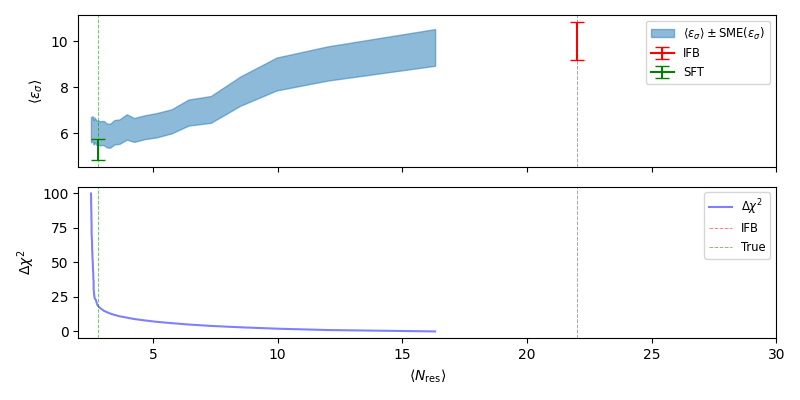}
\caption{Average number of survived resonances $\langle N_{\text{res}} \rangle$ depending on the $\Delta\chi^2$ applied and corresponding error metric values $\langle \varepsilon_{\sigma} \rangle$.
Vertical lines and errorbars show the average number of resonances and performance metric values for the best possible performance (``starting from true'' mode - green line) and for the result ARIS gets after the initial feature bank solve with a redundant number of resonances (red line)}

    \label{fig:result_avg_N_res_A}
\end{figure}

Fig. \ref{fig:result_avg_N_res_A} additionally shows how the elimination of resonances applying different thresholds changes the value of the introduced metric $\varepsilon_{\sigma}$ and the average number of resonances. 

The median value of the error metric in cross section space $\text{med}(\varepsilon_{\sigma})$ has much lower absolute values depending on the threshold used.  This fact can be explained by the distribution of error metrics values estimated for all processed cases with different thresholds applied and lower sensitivity of the median value to outliers.

Study 2 will also investigate the behavior of the second introduced metric, $\varepsilon_{\text {PSF}}$, as a function of the hyperparameter. This metric was introduced to evaluate and compare the quality of spin group assignment, while this study focuses on the correct average parameters and resonances of a single spin group.

\begin{figure}[H]
    \centering
    \includegraphics[width=\textwidth]{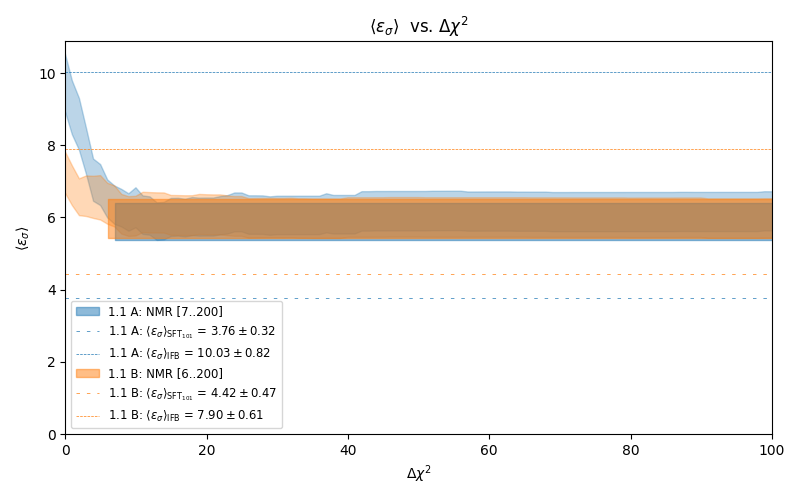}
    \caption{Hyperparameter $\Delta\chi^2$ selection based on the error metric $\langle \varepsilon_{\sigma} \rangle$ (Study 1.1). Horizontal lines indicate values for reference performance estimated using ``starting from true'' results and IFB stage performance (uncertainty values for reference performance are shown in corresponding tables).}
    \label{fig:1.1_AB_avg_e_sigma_comparison}
\end{figure}

To evaluate the effectiveness of the selected hyperparameter on unseen data, dataset B was processed, and the results are presented in Table \ref{tab:res_B_dataset} and Fig. \ref{fig:1.1_AB_avg_e_sigma_comparison}. 

\begin{table}[H]
\centering
\begin{tabular}{l|rr|r}
Metric& \multicolumn{2}{c|}{Estimated}              & Reference       \\
                                  & IFB stage        & Pruning Stage, $\Delta\chi^2_{\text{sel}}$ & SFT$_{101}$     \\ \hline
        $\langle \varepsilon_{\sigma} \rangle $ & $7.9 \pm 0.61$& $6.13 \pm 0.56$& $4.42 \pm 0.47$\\
        med($\varepsilon_{\sigma}$) & $2.01 \pm 0.33$& $1.42 \pm 0.19$& $0.89 \pm 0.11$\\
        $\langle \varepsilon_{\text{PSF}} \rangle $$\times 10^{9}$ & $1.29 \pm 0.15$& $1.81 \pm 0.18$& $1.76 \pm 0.17$\\
        med($\varepsilon_{\text{PSF}}) $$\times 10^{10}$ & $3.31 \pm 0.45$& $2.83 \pm 0.65$& $2.52 \pm 0.45$\\
        $\langle N_{\text{res}} \rangle $ & $19.37 \pm 0.19$& $3.18 \pm 0.05$& $2.74 \pm 0.05$\\
    \end{tabular}
     \caption{Estimated and reference error metrics (Results for Study 1.1, dataset B)}
\label{tab:res_B_dataset}
\end{table}
The performance  estimated using dataset B processing results is consistent with that for dataset A, which suggests that the selected hyperparameter is stable and suitable for data with the same average parameters and assumptions.

\subsection{Influence of Incomplete Data Covariance Matrix and Incorrect Average Parameters}

\begin{figure}[H]
    \centering
    \includegraphics[width=\textwidth]{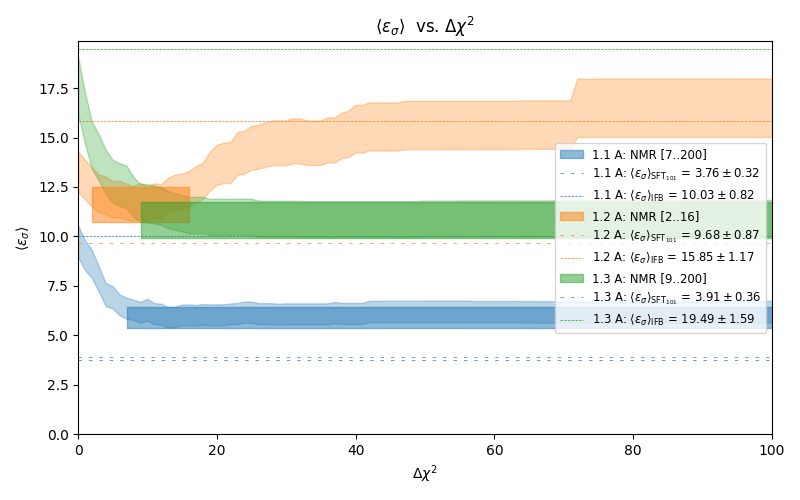}
    \caption{Performance comparison of $\langle \varepsilon_{\sigma} \rangle$ under scenarios where ARIS is supplied with either misreported covariance data (Study 1.2) or incorrect average resonance parameters (Study 1.3). For reference, performance without any distortions using the same dataset is shown (Study 1.1). Horizontal lines represent reference values, including performance using "starting from true" resonance parameters and post-IFB stage performance. Shaded areas indicate the range where the performance metric $\langle \varepsilon_{\sigma}\rangle$ falls within one standard error of the mean around the minimum $\langle \varepsilon_{\sigma} \rangle_{\text{min}} \pm \text{SEM}(\langle \varepsilon_{\sigma} \rangle_{\text{min}})$.}
    \label{fig:1.1_1.3_A_avg_e_sigma_comparison}
\end{figure}

Studies 1.2 and 1.3 explore ARIS's performance under partially distorted data inputs. The results from processing with an incomplete data covariance matrix and for scenarios using incorrect average capture widths are presented in Table \ref{tab:res_1.2_1.3_combined}. For both studies same dataset was used and threshold value determined during Study 1.1 applied. These studies quantify the impact of potential data deficiencies demonstrating the ``what if'' testing capabilities of the framework under development. The comparative analysis is depicted in Fig. \ref{fig:1.1_1.3_A_avg_e_sigma_comparison}.

Near-minimum regions for all presented scenarios that used data with resonances of a single spin group (SSG) overlap in the region $\Delta\chi^2_{\text{SSG}} = 9..16 $. 

\begin{table}[H]
\centering
\begin{tabular}{l|rr|r}
Metric& \multicolumn{2}{c|}{Estimated}              & Reference\\
                                  & IFB stage        & Pruning Stage, $\Delta\chi^2_{\text{sel}}$ & SFT$_{101}$ \\ \toprule
\multicolumn{4}{c}{Study 1.2 (incomplete DCM), dataset A}  \\ \midrule
$\langle \varepsilon_{\sigma} \rangle $ & $15.85 \pm 1.17$& $12.05 \pm 0.88$& $9.68 \pm 0.87$\\
med($\varepsilon_{\sigma}$) & $5.38 \pm 0.87$& $3.15 \pm 0.48$& $1.41 \pm 0.27$\\
$\langle \varepsilon_{\text{PSF}} \rangle$$\times 10^{9}$ & $5.68 \pm 0.44$& $5.4 \pm 0.49$& $6.13 \pm 0.61$\\
med($\varepsilon_{\text{PSF}})$$\times 10^{10}$ & $19.78 \pm 1.42$& $12.44 \pm 2.03$& $4.95 \pm 1.12$\\
$\langle N_{\text{res}} \rangle $ & $26.67 \pm 0.23$& $3.58 \pm 0.07$& $2.79 \pm 0.04$\\ \midrule
\multicolumn{4}{c}{Study 1.3 (incorrect average capture width),  dataset A}  \\ \midrule
$\langle \varepsilon_{\sigma} \rangle $ & $19.49 \pm 1.59$& $11.33 \pm 0.94$& $3.76 \pm 0.32$\\
med($\varepsilon_{\sigma}$) & $4.89 \pm 0.62$& $2.8 \pm 0.41$& $0.91 \pm 0.11$\\
$\langle \varepsilon_{\text{PSF}} \rangle$$\times 10^{10}$ & $10.72 \pm 1.03$& $9.37 \pm 1.21$& $19.11 \pm 2.03$\\
med($\varepsilon_{\text{PSF}})$$\times 10^{10}$ & $3.55 \pm 0.31$& $1.34 \pm 0.2$& $2.45 \pm 0.64$\\
$\langle N_{\text{res}} \rangle $ & $22.18 \pm 0.2$& $3.18 \pm 0.05$& $2.79 \pm 0.04$\\ \bottomrule
\end{tabular}
\caption{Estimated and reference error metrics for Studies 1.2 and 1.3, dataset A}
\label{tab:res_1.2_1.3_combined}
\end{table}

Analyzing the results for performance metric $\langle \varepsilon_{\sigma} \rangle$ calculated at the selected threshold $\chi^2_{\text{sel}}$, it was shown that in scenarios where ARIS was provided with incomplete data covariance matrices, this error metric increased by a factor of 2.05 compared to the Study 1.1 results, which used a complete data covariance matrix. For the same case, but analyzing results for ``start from true'' resonance parameters mode (SFT) - the difference increased by a factor of 2.57. A comparable increase, by a factor of 1.92, was observed for Study 1.3, where ARIS was supplied with incorrect average resonance parameters values and was not allowed to fit capture widths.  

\subsection {Performance comparison (multiple spin groups)}

\begin{figure}[H]
    \centering
    \includegraphics[width=\textwidth]{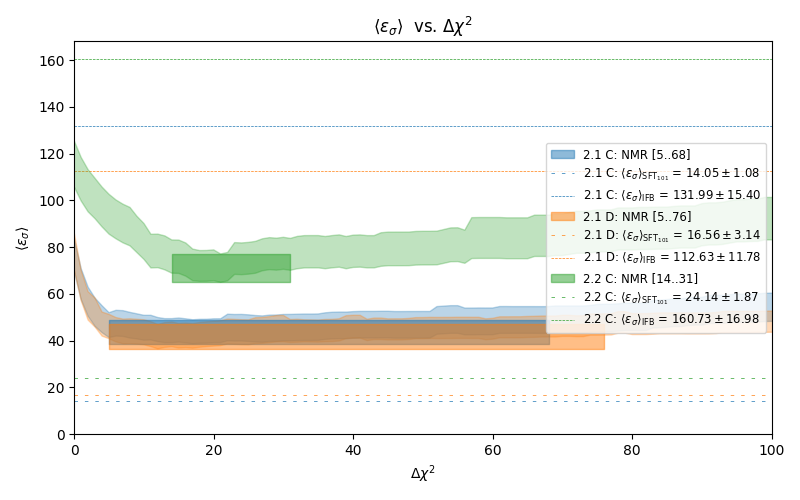}
    \caption{Comparison of ARIS performance calculated as the average error metric $\langle \varepsilon_{\sigma} \rangle$ with a full data covariance matrix (Study 2.1) and incomplete covariance data (Study 2.2). Horizontal lines represent benchmark performances, including scenarios starting from true resonance parameters and post-IFB stage outcomes. Shaded areas show the range where $\langle \varepsilon_{\sigma}\rangle$ lies within one standard error of the mean around the identified minimal error.}
    \label{fig:2.1_CD_avg_e_sigma_comparison}
\end{figure}

Datasets C and D contain resonances of multiple spin groups and were used for threshold selection and validation, respectively (Study 2.1).  Dataset C was also utilized when ARIS was supplied with an incomplete data covariance matrix (Study 2.2).  The comparison of results for estimated averaged error in cross section space is shown in Fig \ref{fig:2.1_CD_avg_e_sigma_comparison}  and in Table \ref{tab:res_study_2_combined}. 

As expected, comparison of the results presented in Tables \ref{tab:res_study_2_combined} and \ref{tab:res_1.1_A} shows that all average error metrics increased significantly for the case of multiple spin groups.

For Study 2.1, when ARIS is supplied with correct data covariance matrices, the performance level is similar, and the near-minimum regions have a significant overlap for processing results for both datasets. However, when ARIS is supplied with an incomplete data covariance matrix, the performance curve exhibits a bowl-like shape, with a near-minimum region around \(\Delta\chi^2_{\text{NMR}_{2.2}} = 14 \ldots 31\).

\begin{table}[H]
\centering
\begin{tabular}{l|rr|r}
Metric& \multicolumn{2}{c|}{Estimated}              & Reference\\
                                  & IFB stage        & Pruning Stage, $\Delta\chi^2_{\text{sel}}$ & SFT$_{101}$ \\ \toprule
\multicolumn{4}{c}{Study 2.1, dataset C}  \\ \midrule
$\langle \varepsilon_{\sigma} \rangle $ & $131.99 \pm 15.4$ & $44.28 \pm 5.14$ & $14.05 \pm 1.08$ \\
med($\varepsilon_{\sigma}$)       & $31.88 \pm 2.75$  & $12.86 \pm 1.26$  & $5.52 \pm 0.58$   \\
$\langle \varepsilon_{\text{SF}} \rangle$$\times 10^{9}$ & $9.7 \pm 1.83$   & $13.88 \pm 1.52$ & $8.65 \pm 0.64$  \\
$\langle \varepsilon_{\text{PSF}} \rangle$$\times 10^{9}$ & $2126.24 \pm 195.64$ & $1259.6 \pm 169.93$ & $6.77 \pm 0.53$ \\
med($\varepsilon_{\text{PSF}})$$\times 10^{9}$ & $373.62 \pm 67.52$ & $51.97 \pm 8.25$  & $2.56 \pm 0.3$    \\
$\langle N_{\text{res}} \rangle $ & $57.98 \pm 0.87$  & $5.95 \pm 0.09$   & $5.77 \pm 0.06$   \\ \midrule
\multicolumn{4}{c}{Study 2.1, dataset D}  \\ \midrule
$\langle \varepsilon_{\sigma} \rangle $ & $112.63 \pm 11.78$ & $42.43 \pm 5.3$  & $16.56 \pm 3.14$ \\
med($\varepsilon_{\sigma}$)       & $32.76 \pm 2.69$  & $11.27 \pm 0.95$  & $4.52 \pm 0.48$   \\
$\langle \varepsilon_{\text{SF}} \rangle$$\times 10^{9}$ & $11.14 \pm 1.9$   & $12.45 \pm 1.91$ & $7.34 \pm 0.85$  \\
$\langle \varepsilon_{\text{PSF}} \rangle$$\times 10^{8}$ & $219.59 \pm 22.61$ & $163.22 \pm 21.96$ & $1.18 \pm 0.56$ \\
med($\varepsilon_{\text{PSF}})$$\times 10^{9}$ & $262.89 \pm 69.16$ & $59.51 \pm 14.06$  & $1.87 \pm 0.16$   \\
$\langle N_{\text{res}} \rangle $ & $53.9 \pm 0.82$   & $6.09 \pm 0.09$   & $5.72 \pm 0.06$   \\ \midrule
\multicolumn{4}{c}{Study 2.2, dataset C}  \\ \midrule
$\langle \varepsilon_{\sigma} \rangle $ & $160.73 \pm 16.98$& $77.57 \pm 7.26$& $24.14 \pm 1.87$ \\
med($\varepsilon_{\sigma}$) & $43.23 \pm 5.43$& $27.49 \pm 2.05$& $7.71 \pm 0.89$ \\
$\langle \varepsilon_{\text{SF}} \rangle$$\times 10^{8}$ & $1.98 \pm 0.24$& $2.69 \pm 0.28$& $2.13 \pm 0.18$ \\
$\langle \varepsilon_{\text{PSF}} \rangle $$\times 10^{8}$ & $227.38 \pm 25.42$& $185.35 \pm 22.25$& $1.4 \pm 0.11$ \\
med($\varepsilon_{\text{PSF}}) $$\times 10^{9}$ & $345.28 \pm 69.85$& $154.43 \pm 18.3$& $3.46 \pm 0.63$ \\
$\langle N_{\text{res}} \rangle $ & $64.72 \pm 0.89$& $7.27 \pm 0.14$& $5.77 \pm 0.06$ \\
\bottomrule
\end{tabular}
\caption{Estimated and reference performance for Studies 2.1 and 2.2}
\label{tab:res_study_2_combined}
\end{table}

\begin{figure}[H]
    \centering
    \includegraphics[width=\textwidth]{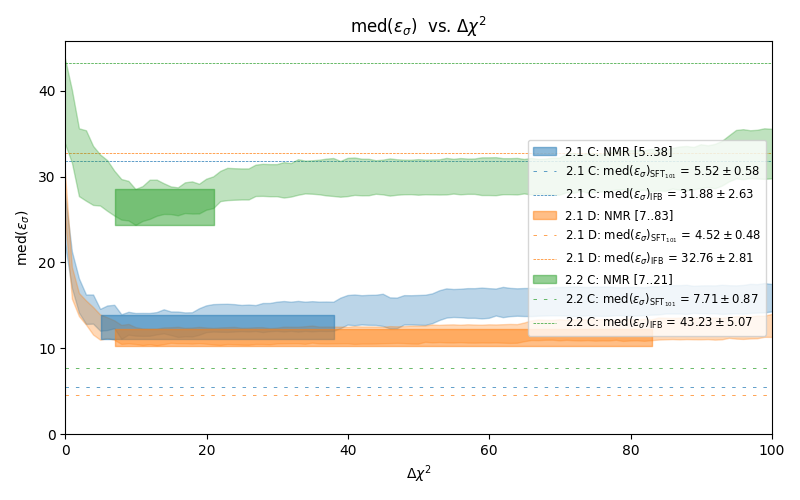}
    \caption{Comparison of ARIS performance utilizing the estimated median of error metric $\text{med} ( \varepsilon_{\sigma} ) $ with a full data covariance matrix (Study 2.1) versus incomplete covariance data (Study 2.2).} 
    \label{fig:2.1_CD_med_e_sigma_comparison}
\end{figure}

The median values of the error metric are almost three times lower than the calculated mean values and behave slightly differently depending on the threshold value, assuming some ``outlier cases" that significantly impact the average value - see Fig. \ref{fig:2.1_CD_med_e_sigma_comparison}.  

Additional constraints can be incorporated into the analysis such that the estimates for the average and median values for both metrics are minimized. For Group 2 studies, the suboptimal threshold values lie in \(\Delta\chi^2_{\text{MSG}} = 13 \ldots 21\).

If we compare regions of similar performance levels for single and multiple spin groups, it can be observed that the results for both studies overlap in the region \(13 \ldots 16\). The position of this region in the presented studies is primarily driven by the results of the tests with incomplete data covariance matrices (Studies 1.2 and 2.2), where the calculated performance curves exhibit clear minima. Excluding those cases and considering only the results from studies with correct data covariance matrices and/or incorrect average parameters supplied to ARIS, the estimated region of ARIS's suboptimal performance was universally set to be \(\Delta\chi^2_{\text{SO}} = 9 \ldots 38\).

In our case, for all the results shown in the next tables, \(\Delta\chi_{\text{sel}}^2 = 13 \) is used since it is the most conservative value: the smaller the value of the hyperparameter, the more resonances survive the pruning stage, but at the same time the number of ``fake" resonances also increases. This threshold value also corresponds to the minimum value of the weighted average error for $\langle \varepsilon_{\sigma} \rangle$ across all described scenarios (assuming all of them are equally probable).

The comparison of processing results reveals that a hyperparameter optimized for a dataset with resonances from a single spin group effectively improves a selected error metric in datasets with resonances from multiple spin groups. This can be attributed to the similar resonance parameters and almost identical signal-to-noise ratios for both studies. 

\begin{figure}[H]
\centering
\includegraphics[width=\textwidth]{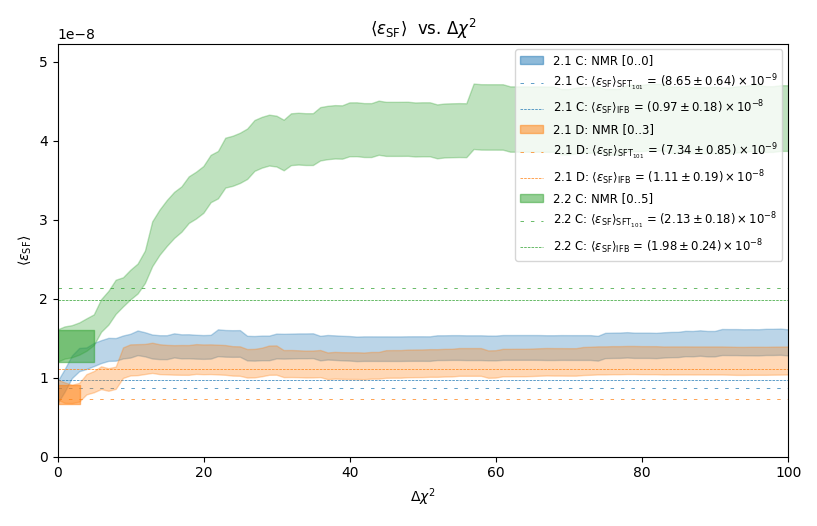}
\caption{Comparison of ARIS performance using the estimated averaged error metric value $\langle \varepsilon_{\text{SF}} \rangle$ with a full data covariance matrix versus incomplete covariance data.}
\label{fig:2.1_CD_avg_e_SF_comparison}
\end{figure}

\begin{figure}[H]
\centering
\includegraphics[width=\textwidth]{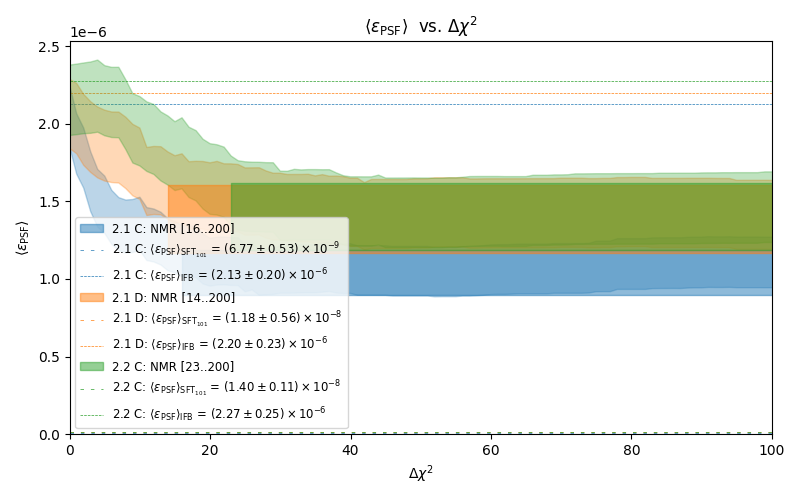}
\caption{Comparison of ARIS performance using the estimated averaged error metric value $\langle \varepsilon_{\text{PSF}} \rangle$ with a full data covariance matrix versus incomplete covariance data.}
\label{fig:2.1_CD_avg_e_PSF_comparison}
\end{figure}

Comparing the behavior of the metrics based on the strength functions, we can see significant differences depending on the hyperparameter value - see Fig. \ref{fig:2.1_CD_avg_e_SF_comparison}, \ref{fig:2.1_CD_avg_e_PSF_comparison}. Specifically, the metric based on the strength function $\langle \varepsilon_{\text{SF}} \rangle$ shows results for cases with complete and correct covariance data that are close to, or even better than, reference values calculated for SFT mode. Thus ARIS with current configuration, on average, performs only 1.7 times worse than when true resonance parameters are used as a prior. This result is believed to be due to an excessive number of small resonances retained after the IFB stage, which align well with many small neutron widths for resonances of all spin groups while using a fixed capture width mode.

In contrast, for the worst-case scenario, the error metric $\langle \varepsilon_{\text{PSF}} \rangle$ is approximately 187 times larger compared to results when using true resonance parameters as a prior. Such differences can be explained by the fact that during the IFB stage, the data was fitted without any considerations on spin group assignment, allowing all potential resonances equal chances of being selected, yet only a few were retained based on the internal logic of the fitting process. Furthermore, since all fitting stages utilized only $\chi^2$ as the objective function without additional constraints, the spin groups of surviving resonances remained unchanged while their parameters were adjusted to fit the data. During the pruning stage, the number of resonances was reduced, resulting in variations in neutron width values while capture width values remained fixed. For this test case, applying the selected hyperparameter value increased ARIS performance by 23\% even without spin group reassignment (for the worst-case scenario, compared with the IFB stage results). Notably, resonances from both spin groups have similar average capture widths for a given isotope. If there were significant differences in capture widths between spin groups, errors in spin group assignment combined with fixed capture width settings during fitting could potentially decrease performance, necessitating further investigation. This scenario raises important questions for further research: the impact of spin group assignment on the results for both $\langle \varepsilon_{\text{PSF}} \rangle$ and $\langle \varepsilon_{\sigma} \rangle$ metrics and investigation on how different spin group selection techniques applied at different stages of fitting might affect processing outcomes (benchmarking of the spin group selection techniques utilizing introduced metrics).

These results exemplify how synthetic data can be used for process optimization and hypothesis testing. They highlight the sensitivity of all introduced metrics to the hyperparameter value and underline the importance of understanding the underlying processes that influence parameter adjustment.

\section{Conclusions}

This research demonstrates the effective application of high-fidelity synthetic data in automating key parts of the nuclear data evaluation process. Through the utilization of high-fidelity synthetic data and the introduction of key metrics of interest, a methodology is established that improves the precision, reproducibility, and objectivity of any investigated process. This approach helps to mitigate the subjective biases inherent in traditional manual evaluation methods, resulting in a more systematic and efficient process. The cornerstone of this methodology is the ability to reference a true solution, which, combined with the definition of several quantitative metrics, enables comprehensive analyses for benchmarking and optimizing a nuclear data evaluation.  

In particular, two quantitative error metrics were introduced to assess and optimize the performance of the automated resonance identification process and corresponding subroutine. The ability to select hyperparameter values was demonstrated. A comparison of the subroutine's functionality before and after optimization demonstrates the efficacy of the developed approach in terms of the defined error metrics. 

It should be noted that the hyperparameter values selected deviate from those expected theoretically because of the violated assumptions (linear models,  Gaussian distributions). Despite this, the chosen values enhance performance, demonstrating that this technique is effective for selecting hyperparameters. 

Synthetic data enabled testing of the impact of inaccuracies in the data covariance matrix or incorrect average resonance parameters on resonance identification accuracy and performance using introduced metrics to compare results in cross-section space. The results also show the robustness of this approach: even when the ARIS gets distorted data covariance matrices or significantly biased values for average resonance parameters the selected hyperparameters decrease the selected error metric values. 

The assessment results highlight that accurate spin group assignment is crucial for improving the performance of both metrics introduced, significantly impacting the efficacy and mechanics of the identification and fitting processes. Therefore, investigating more sophisticated methods of spin group assignment is essential, and further comparative analyses are required to refine this aspect of the automated fitting methodology used. 

The focus must also be shifted towards study on the importance of accurate spin group assignment, development and application of corresponding techniques to enhance overall performance and processing efficiency. Additionally, a thorough examination of the fitting technique and developed automated resonance identification tool across various energy ranges and assumptions is required to formulate recommendations to increase accuracy and stability and make this tool more robust. Another prospective question is the benchmarking of the resonance identification tools testing different assumptions about the data and utilizing prior knowledge (applying Bayesian inference), which will potentially increase the accuracy and simplify the procedure. 

Furthermore, synthetic data enables the testing of various hypotheses related to underlying physical models and corresponding assumptions. This includes evaluator decisions such as where to stop the resolved resonance evaluation and theoretical assumptions about parameter distributions. Future efforts will also extend to uncertainty quantification and validation. The controlled environment provided by this framework offers substantial opportunities for simulating experiments and refining experimental designs, thereby enhancing overall methodology and outcomes.

\newpage
\section*{Acknowledgements}
This material is based upon work supported by the Department of Energy National Nuclear Security Administration through the Nuclear Science and Security Consortium under Award Number(s) DE-NA0003996.

This report was prepared by the research group of Dr. Sobes partially under award 31310021M0041 from Assistance Agreements, Nuclear Regulatory Commission. The statements, findings, conclusions, and recommendations are those of the author(s) and do not necessarily reflect the view of the Assistance Agreements or the US Nuclear Regulatory Commission.

This work was supported by the Nuclear Criticality Safety Program, funded and managed by the National Nuclear Security Administration for the Department of Energy. This report was prepared as an account of work sponsored by an agency of the United States Government. Neither the United States Government nor any agency thereof, nor any of their employees, makes any warranty, express or implied, or assumes any legal liability or responsibility for the accuracy, completeness, or usefulness of any information, apparatus, product, or process disclosed, or represents that its use would not infringe privately owned rights. Reference herein to any specific commercial product, process, or service by trade name, trademark, manufacturer, or otherwise does not necessarily constitute or imply its endorsement, recommendation, or favoring by the United States Government or any agency thereof. The views and opinions of authors expressed herein do not necessarily state or reflect those of the United States Government or any agency thereof.

\section*{Declaration of generative AI and AI-assisted technologies in the writing process}

During the preparation of this work, the authors used ChatGPT for language enhancement. After using this tool, the authors reviewed and edited the content as needed and take full responsibility for the publication's content.

 \bibliographystyle{elsarticle-num} 
 \bibliography{cas-refs}





\end{document}